\documentclass{svproc}
\usepackage{amsmath, amssymb}
\usepackage{algpseudocode,algorithm}
\usepackage{graphicx}
\usepackage{subcaption}
\usepackage{hyperref}
\makeatletter
\newcommand{\printfnsymbol}[1]{%
	\textsuperscript{\@fnsymbol{#1}}%
}
\makeatother
\makeatletter
\newcommand{\mathleft}{\@fleqntrue\@mathmargin0pt}
\newcommand{\mathcenter}{\@fleqnfalse}
\makeatother
\usepackage{url}

\usepackage{caption}
\usepackage{subcaption}
\usepackage{cite}
\usepackage{graphicx}
\usepackage{textcomp}
\usepackage{xcolor}
\usepackage{algorithm}
\usepackage{tabularx}
\usepackage{adjustbox}
\usepackage{float}
\usepackage{placeins}
\usepackage{booktabs}

\begin{document}
	\mainmatter              
	
	\title{Nondiagonal Mixture of Dirichlet Network Distributions for Analyzing a Stock Ownership Network}
	\titlerunning{Nondiagonal Mixture of Dirichlet Network Distributions} 

\author{Wenning Zhang\inst{1} \and Ryohei Hisano\inst{1,4}\thanks{Corresponding author.  The first and second authors made equal contributions.}
	Takaaki Ohnishi\inst{2,4} \and Takayuki Mizuno\inst{3,4}}
\authorrunning{Zhang et al.} 
%
%
\institute{Graduate School of Information Science and Technology, the University of Tokyo\\
	\and
	Graduate School of Artificial Intelligence and Science, Rikkyo University
	\and
	National Institute of Informatics
	\and
	The Canon Institute for Global Studies
}

	\maketitle              

\begin{abstract}
	Block modeling is widely used in studies on complex networks. The cornerstone model is the stochastic block model (SBM), widely used over the past decades. However, the SBM is limited in analyzing complex networks as the model is, in essence, a random graph model that cannot reproduce the basic properties of many complex networks, such as sparsity and heavy-tailed degree distribution. In this paper, we provide an edge exchangeable block model that incorporates such basic features and simultaneously infers the latent block structure of a given complex network. Our model is a Bayesian nonparametric model that flexibly estimates the number of blocks and takes into account the possibility of unseen nodes. Using one synthetic dataset and one real-world stock ownership dataset, we show that our model outperforms state-of-the-art SBMs for held-out link prediction tasks.
\end{abstract}

\section{Introduction}
Block modeling has been widely used in studies on complex networks\cite{Newman2010,Doreian2019}. The goal of block modeling is to uncover the latent group memberships of nodes responsible for generating the complex network. The uncovered latent block structure is used for both prediction and interpretation.  For prediction, block modeling is used to find missing or spurious edges\cite{Liben-Nowell2007,Martinez2016}.  For interpretation, the estimated latent block structure provides a coarse-grained summary of the linkage structure that is particularly useful in complex networks, which is often messy at the primary level.

The cornerstone model of block modeling is the stochastic block model (SBM)\cite{Holland1981,Snijders1997,Nowicki2001}. In the SBM, each node is assigned to a block.  The edge probability in the network is governed solely by the linkage probability defined among these blocks.  The goal of the SBM is to find the latent block structure and the linkage probability among the blocks. If given only one block structure, the model collapses to the Erdős–Rényi--Gilbert type random graph model that dates back to the 1950s\cite{Erdos1959,Bollobas2001}.

The fact that the random graph model cannot reproduce basic properties, such as the sparsity and heavy-tailed degree distribution of complex networks, has always been an issue\cite{Barabasi1999,Newman2010}.  The failure of random graph models to reproduce these basic properties has recently been re-examined from the perspective of node exchangeability \cite{Crane2018}.  From the graphon formulation \cite{Bickel2009} and Aldous--Hoover theorem\cite{Aldous1981,Hoover1979}, it can be proven that the only possible network in the random graph model setting is either dense or empty\cite{CraneDempsey2018,Crane2018}.  This limitation makes the SBM rather unsuitable for modeling complex linkage patterns found in various complex networks.

Several authors have proposed models that go beyond such limitations using these modern findings.  One line of research uses exchangeable point processes to generate the linkage patterns in a network\cite{CaronFox2017}.  In their formulation, edges appear when a pair of nodes occur in a nearby time position in the point processes.  \cite{CaronFox2017} showed that this formulation could generate sparse networks. Another line of research focuses on a more intuitive edge generation process based on edge exchangeability\cite{Crane2018,CraneDempsey2018,Cai2016,CraneHollywood2017}.  Edge exchangeable models have been proven to generate a sparse and heavy-tailed network. \cite{Williamson2016} proposed a model that considers the latent community structure in addition to the edge exchangeable framework.  They called their model the mixture of Dirichlet network distributions (MDND)\cite{Williamson2016}.  

However, the MDND oversimplifies the latent block structure limiting it to only the diagonal case, similar to community detection algorithms.  These limitations are problematic in instances in which the flow of influences (or information) among the blocks is the focus of research.  One such example is the stock ownership network.  In this setting, companies consider direct ownership and indirect ownership to maximize their influence and minimize risks\cite{Garcia2017}.  A simple diagonal block structure only provides community-like clustering of companies, which is unsatisfactory.

In this paper, we provide a nondiagonal extension of the MDND (the NDMDND model) that makes it possible to estimate both the diagonal and nondiagonal latent block structure.  Our model has no additional limitations than the MDND, and flexibly infers the number of blocks and considers the possibility of unseen nodes.  It is noteworthy that our model can be regarded as a nonparametric extension of the sparse block model\cite{Parkkinen2009}.  The sparse block model is a precursor model that focused on edge exchangeability even before the connection between sparse graphs and edge exchangeability was rigorously proven.  We highlight both models in this paper.


\section{Related works}
\subsection{Sparse block model}\label{sec:SparseBlock}
In this section, we provide a brief explanation of the sparse block model. We use the notation $(s_n,r_n)$ to denote the $n$th edge of the network, and $c_n=(c_{sn},c_{rn})$ to denote the block pair to which that $n$th edge is assigned.  We use $A_{k}$ to define the node proportion distribution that captures which nodes are probable in block $k$.  We use $Dir()$ to denote the Dirichlet distribution and $Cat()$ the categorical distribution, where the parameters are written inside the parentheses.  We summarize the generative process as follows:
\begin{tabbing}
	\hspace{10pt}\=\hspace{10pt}\=\kill
	\textbf{(A) Initialization} \\
	\> \textbf{For} each block pair $k = 1 , \ldots , K$, \\
	\>\>  we draw the node proportions $A_k\sim\mbox{Dir}(\tau)$ \\
	\textbf{(B) Sampling of block pairs and edges} \\
	\>\textbf{For} each edge $(s_{n},r_{n})$, \\
	\>\> (1) sample the block pair $c_{n}=(c_{sn},c_{rn}) \sim\mbox{Cat}(\theta)$ \\
	\>\> (2) sample the sender node from $s_n\sim\mbox{Cat}(A_{c_{sn}})$ \\
	\>\> (3) sample the receiver node from $r_n\sim\mbox{Cat}(A_{c_{rn}})$.
\end{tabbing}

Note that in the sparse block model, the latent block structure is defined in advance.  The goal of the sparse block model is to infer the probability of each block to generate nodes (i.e., $A_{k}$), and the probability of each block pair (i.e., $c_{n}$) appearing from a given network.  The fact that we have to specify the latent block structure is a huge limitation.  It implies that we have to provide both the number of blocks to use and the block pairs' interaction patterns before seeing the data.  Second, note that the same node pairs could appear multiple times in this setting (i.e., multigraph).  These multiple edges could be used as a proxy for the edge weights.  Although we could add a link function that links the proxy edge weights to the continuous edge weights, in this paper, we make the simple assumption that these multiple occurrences of an edge describe the weights of an edge. Finally, note also that the number of nodes used in the network is fixed; it does not increase as we sample more edges in the process.

\subsection{Mixture of Dirichlet network distributions}
The MDND is a nonparametric Bayesian model that attempts to infer the number of blocks from the observed network. Using a Bayesian formulation, it is also possible to estimate the probability of unseen nodes in sharp contrast to the sparse block model. The MDND assumes a diagonal block structure for the latent block structure and uses the Chinese restaurant process\cite{Phadia2013} to model the diagonal block pair linkage probability.  The modeling of the probability of nodes given a block is more involved.  Assume that a Chinese restaurant process for each block leads to each block's own set of nodes.  For the model to force all the blocks to use the same set of nodes, we need to extend the Chinese restaurant process to the Chinese restaurant franchise process\cite{Teh2006}.  The Chinese restaurant franchise process introduces an auxiliary assignment variable called a ``table''.  By separating the growth of the popularity of tables and the assignment of nodes (i.e., in \cite{Teh2006}s term ``dish'') to the table, we can make multiple Chinese restaurant processes share the same set of nodes.
We use $CRP(\alpha)$ to denote the Chinese restaurant process with hyperparameter $\alpha$. We use subscripts to discern the multiple Chinese restaurant processes used in the model.  We use $s_{nt}$ and $r_{nt}$ to denote the table assigned to the sending node that originates from the Chinese restaurant franchise process. $\alpha$, $\tau$, and $\gamma$ are hyperparameters of the model. The generative process is as follows:

\begin{tabbing}
	\hspace{10pt}\=\hspace{10pt}\=\kill
	\textbf{(A) Sampling of diagonal blocks} \\
	\>\textbf{For} each edge sample $c_n \sim CRP_{B}(\alpha)$ where $c_{sn}$ is always equal to $c_{rn}$\\
	\textbf{(B) Sampling of edges} \\
	\>(1) Sample a table for the sender node: $s_{nt} \sim CRP_{c_{n}}(\gamma)$ \\
	\>\> \textbf{if} $s_{nt}$ is a new table, then sample $s_{n} \sim CRP_{N}(\gamma)$ \\
	\>\> \textbf{else} $s_{n}$ is assigned the same node as $s_{nt}$ \\
	\>(2) Sample a table for the receiver node:  $r_{nt} \sim CRP_{c_{n}}(\gamma)$ \\
	\>\> \textbf{if} $r_{nt}$ is a new table, then sample $r_{n} \sim CRP_{N}(\gamma)$ \\
	\>\> \textbf{else} $r_{n}$ is assigned the same node as $r_{nt}$.
\end{tabbing}

\section{Nondiagonal mixture of Dirichlet network distributions}
\subsection{Generating process}
Our proposed model, the NDMDND, can be considered as both a nonparametric Bayesian counterpart of the sparse block model and a nondiagonal extension of the MDND.  Our model can be created by adding two components to the MDND: (1) adding another Chinese restaurant process that controls the number of block pairs used to model the latent block structure and (2) modifying the Chinese restaurant process that governs the appearance of blocks in the MDND to the Chinese restaurant franchise process.  The latter extension is necessary because, as in the node-set case in the MDND, assuming a Chinese restaurant process separately for the sender blocks and receiver blocks would lead to each side having its own set of blocks.  To prevent this, we need to make sure that both the sender and receiver sides share the same set of blocks.  The node generation mechanism could be the same as in the MDND case without any further extension.

In the MDND, we need to add four Chinese restaurant processes: one for the block pair table (which we denote as $CRP_{block-pair}(\tau_{pair})$), one for the block tables for the sending nodes ($CRP_{block-send}(\tau_{block})$), one for the block tables for the receiving nodes ($CRP_{block-rece}(\tau_{block})$), and the last one responsible for generating the new blocks ($CRP_{block}(\gamma_{block})$).  We use $c_{nt} = (c_{snt},c_{rnt})$ to denote the pair table assigned to each edge.  We further use $s_{nbt}$ and $r_{nbt}$ to denote the block tables assigned to the sender and receiver nodes, and $s_{nb}$ and $r_{nb}$ to denote the block assigned to each node.  The generative process is as follows:

\begin{tabbing}
	\hspace{10pt}\=\hspace{10pt}\=\hspace{10pt}\=\kill
	\textbf{(A) Sampling of block pairs} \\
	\> \textbf{For} each edge sample pair table $c_{nt} \sim CRP_{block-pair}(\tau_{pair})$ \\
	\> \textbf{if} $c_{nt}$ is a new pair table \\
	\>\> (1) Sample $s_{nbt} \sim CRP_{block-send}(\tau_{block})$ :\\
	\>\>  \textbf{if} $s_{nbt}$ is a new send block table, then sample $s_{nb} \sim CRP_{block}(\gamma_{block}) $ \\
	\>\> \textbf{else} assign the block associated to $s_{nbt}$ to $s_{nb}$ \\
	
	\>\> (2) Sample $r_{nbt} \sim CRP_{block-rece}(\tau_{block})$: \\
	\>\> \textbf{if} $r_{nbt}$ is a new send block table sample $r_{nb} \sim CRP_{block}(\gamma_{block}) $ \\
	\>\>  \textbf{else} assign the block associated to $r_{nbt}$ to $r_{nb}$ \\
	\> \textbf{else} assign the block table and block pair associated to the $c_{nt}$ to\\
	\>\>\>  $(s_{nbt},r_{nbt})$ and $(s_{nb},r_{nb})$ \\
	\textbf{(B) Sampling of edges} \\
	\>(1) Sample a table for the sender node: $s_{nt} \sim CRP_{c_{n}}(\gamma)$ \\
	\>\> \textbf{if} $s_{nt}$ is a new table then sample $s_{n} \sim CRP_{N}(\gamma)$ \\
	\>\> \textbf{else} $s_{n}$ is assigned the same node as $s_{nt}$ \\
	\>(2) Sample a table for the receiver node:  $r_{nt} \sim CRP_{c_{n}}(\gamma)$ \\
	\>\> \textbf{if} $r_{nt}$ is a new table, then sample $r_{n} \sim CRP_{N}(\gamma)$ \\
	\>\> \textbf{else} $r_{n}$ is assigned the same node as $r_{nt}$ 
\end{tabbing}

In NDMDND, $\gamma_{block}$ controls the number of blocks used.  A low $\gamma_{block}$ with a relatively high $\tau_{pair}$ would lead to a more dense structure, whereas increasing $\gamma_{block}$ would make the number of blocks increase.  $\tau_{pair}$ and $\tau_{block}$ are trickier to interpret as both parameters also affect the possibility of considering a new block or block pair in the model.  We further explain this issue in the next section.

\begin{algorithm}
	\caption{Inference algorithm of NDMDND} 
	\label{ndmdnd}
	\begin{algorithmic}
		\While{not converged}
		\State{Update $\beta$s using $\textbf{t}$ and $\textbf{k}$}
		\For{$q = 1 , \ldots , T_{1}$}
		\State{Sample edge $i$ at random}
		\State{Sample from
			\begin{equation}
				\text{$p(t_{p}^{i}=t_{p}|t_{p}^{-i},\textbf{k})$} \propto  
				\begin{cases} 
					\text{$n_{t_{p}}^{-i} f_{k_{t_{s}},k_{t_{r}}}^{-s_{i},-r_{i}}(s_{i},r_{i})   $}\\
					\text{$\tau_{p} p(s_{i},r_{i} | t_{p}^{-i},t_{p}^{i}=new,\textbf{k})$}
				\end{cases}
		\end{equation}}
		\If{$\hat{t_{p}^{i}}==new$}
		\State{Sample from 
			\begin{equation}
				\text{$p(t_{s}^{i}=t_{s}|t_{s}^{-i},\textbf{k})$} \propto  
				\begin{cases} 
					\text{$n_{t_{s}}^{-i} f_{k_{t_{s}}}^{-s_{i}}(s_{i})   $}\\
					\text{$\tau_{p} p(s_{i} | t_{s}^{-i},t_{s}^{i}=new,\textbf{k})$}
				\end{cases}
		\end{equation}}
		\If{$ \hat{t_{s}^{i}} ==new$}
		\State{Sample from 
			\begin{equation}
				\text{$p(k_{t_{s}^{i}}=k|\textbf{t},k^{-t_{s}^{i}})$} \propto  
				\begin{cases} 
					\text{$m_{.k}f_{k}^{-s_{i}}(s_{i}) $}\\
					\text{$\gamma_{block} f_{new}^{-s_{i}}(s_{i}) $}
				\end{cases}
		\end{equation}}
		\If{$ \hat{k_{t_{s}^{i}}} ==new$}
		\State{Create a new block and assign the new block to the new table}
		\Else
		\State{Assign $\hat{k_{t_{s}^{i}}}$ to the new table}
		\EndIf
		\Else
		\State{$t_{s}^{i}=\hat{t_{s}^{i}}$}		
		\EndIf
		\State{$\cdots$ Perform exactly the same steps for the receiver blocks}
		\Else
		\State{Assign $t_{p}^{i} = \hat{t_{p}}$ and the accompanying send block table, and rece block }
		\State{ table to $t_{s}^{i}$ and $t_{r}^{i}$}
		\EndIf
		\EndFor
		
		\For{$q = 1 , \ldots , T_{2}$}
		Sample table number $i$ from the sender tables
		\State{Sample from
			\begin{equation}
				\text{$p(k_{t_{s}^{i}}=k|\textbf{t},k^{-t_{s}^{i}})$} \propto  
				\begin{cases} 
					\text{$m_{k}^{-t_{s}^{i}} f_{k}^{-s_{t_{s}^{i}}}(s_{t_{s}^{i}})   $}\\
					\text{$\gamma_{b} f_{k_{new}}^{-s_{t_{s}^{i}}}(s_{t_{s}^{i}})$}
				\end{cases}
		\end{equation}}
		\State{$\cdots$ Perform exactly the same steps for the receiver block tables}
		\EndFor
		\EndWhile
	\end{algorithmic}
\end{algorithm}

\subsection{Inference}

The inference of NDMDND is rather involved compared with that of the MDND counterpart.  In MDND, the direct sampling scheme is used to avoid the sampling of table assignments (Section 5.3 in ~\cite{Teh2006}).  However, in NDMDND, the sampling of both the table and table-to-block assignments turns out to be much simple (Section 5.1 in ~\cite{Teh2006}).  Moreover, a bonus of explicitly sampling tables is that we do not need to simulate the node counts (i.e., the number of tables with block $k$ for a given node $i$, $\rho_{k,i}^{(1)}$ and $\rho_{k,i}^{(2)}$ in \cite{Williamson2016}) and instead evaluate them from our table assignments.  We used these values to estimate the probability of a node appearing in an edge without block pairs.  This probability is defined for both already seen nodes $\beta_1,\cdots,\beta_J$ and unseen nodes  $\beta_{u}$.  A simple sampling relation derives these $ \beta$s: $\beta_1,\cdots,\beta_J,\beta_u\sim\mbox{Dir}(\rho_{\cdot 1}^{(\cdot)},\cdots,\rho_{\cdot J}^{(\cdot)},\gamma)$ where $\rho_{\cdot i}^{\cdot}=\sum_{k}\rho_{k,i}^{(1)}+\rho_{k,i}^{(2)}$ represents the number of tables that a node $i(i\in\{1,\cdots,J,J+1\})$ is selected in all the blocks.



Before introducing the inference scheme in more detail, we need to introduce some further notation. We use $n_{t_{p}}$, $n_{t_{s}}$, and $n_{t_{r}}$ to count the number of edges or nodes assigned to a particular pair block table $t$, send block table $s$, and receive block table $r$, respectively.  We use $n_{t_{p}}^{-i}$ to denote the count, ignoring the $i$th edge.  We sometimes use the subscript $i$ to denote the $i$th table, as in $t_{p}^{i}$, $t_{s}^{i}$, and $k_{t_{s}^{i}}$. Furthermore, we use $m_{k}$ to denote the number of tables associated with block $k$ among both sender block tables and receiver block tables.  Each send and receive block table is associated with a particular block.  We use $k_{t_{s}^{i}}$ and $k_{t_{r}^{i}}$ to denote the block associated with the $i$th send block table and $i$th receive block table, respectively.  We further use $f$ to denote the likelihood and the symbol (i.e., ``$\hat{\,\,\,}$'') to represent the sampled value.

With this additional notation, we can outline the inference algorithm (Algorithm \ref{ndmdnd}). In essence, the algorithm is a composition of the collapsed Gibbs sampling scheme. The first if-else branch considers whether to cluster the new edge to already existing block pair tables or create a new block pair table. If the latter is chosen, we have to consider two cases. One is to use existing block tables to generate the new block pairs, and the other is to create a new table to create the new block pair table. To judge whether to use existing block tables, we separate the sampling into sampling sender block tables and receiver block tables. If a new block table is chosen, we proceed in sampling a block assignment for the table (i.e., $k_{t_{s}^{i}}$ or $k_{t_{r}^{i}}$). The probability of assigning a new block is governed by $\gamma_{block}$.  Algorithm \ref{ndmdnd} makes it clear that setting $\tau_{pair}$ too low would lead to the slow convergence of the MCMC. Therefore, in this paper, we set all the hyperparameters to $\tau_{pair}=100$,$\tau_{block}=10$,$\gamma_{block}=10$,$\tau_{node}= 10$. The modification of the parameters did not change the main result in the paper provided $\tau_{pair}$,$\tau_{block}$,$\gamma_{block}$ was sufficiently high for the sampler to find the correct block structure and sufficiently low for it not to outweigh the likelihood term.

\section{Results}

\subsection{Dataset}

\begin{figure}
	\caption{Degree distribution}
	\label{fig:degree}
	\begin{subfigure}{.49\textwidth}
		\centering
		\includegraphics[width=.7\linewidth]{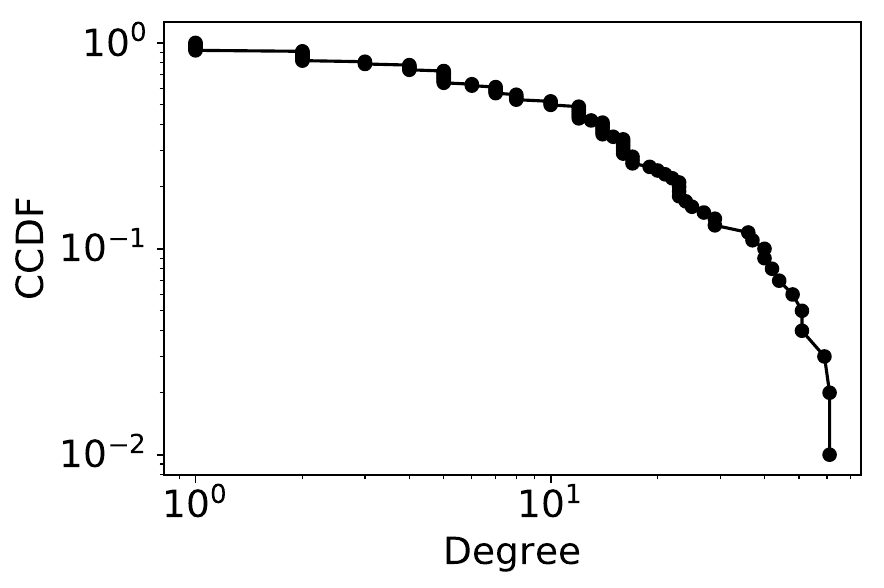}  
		\caption{Synthetic data}
		\label{fig:sub-first}
	\end{subfigure}
	\begin{subfigure}{.49\textwidth}
		\centering
		\includegraphics[width=.7\linewidth]{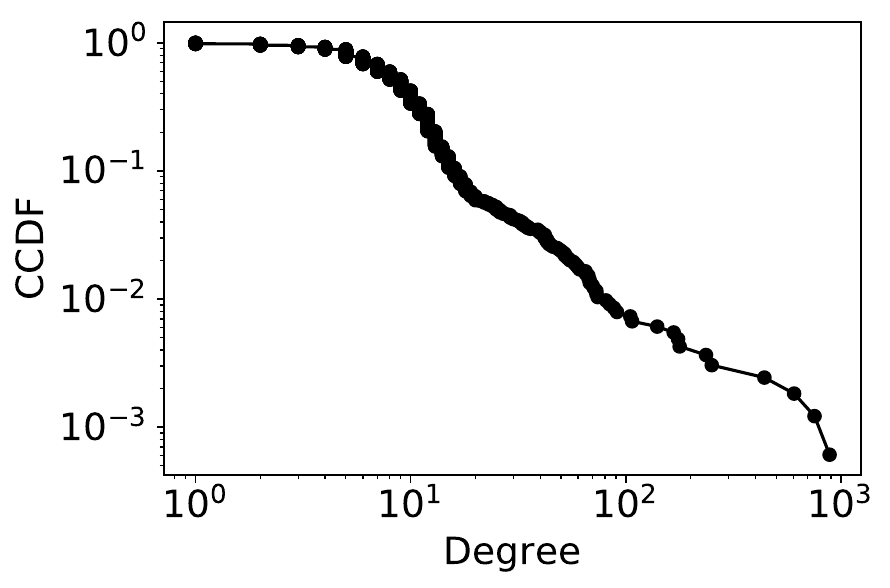}  
		\caption{Stock ownership data}
		\label{fig:sub-first}
	\end{subfigure}
\end{figure}

Our experiments used two datasets: one containing synthetic data and the other containing real-world global stock ownership network data.  The synthetic data was created, assuming the sparse block model.  The stock ownership network is a subset of the Thomson Reuters ownership database.  We focused on the ownership of significant assets in the second quarter of 2015.  Both can be considered as a weighted network, and the datasets' basic summary statistics are shown in Table \ref{data}\footnote{The weights for the stock ownership data is in percentage term.}.  In both datasets, the network is sparse: the synthetic data has 7.2 percent, and the stock ownership data has 0.4 percent of all possible edges.  Moreover, both datasets exhibit a heavy-tailed degree distribution, as shown in Fig.~\ref{fig:degree}.

\begin{table}
	\caption{Datasets}
	\label{data}
	\begin{center}
		\resizebox{0.7\columnwidth}{!}{
			\begin{tabular}{c@{\quad}cccccc}
				\hline
				Dataset & Number of & Number of & Min & Max & Min & Max\\
				&nodes & edges& degree & degree & weight & weight\\[2pt]
				\hline\rule{0pt}{12pt}
				Synthetic &100 & 719 & 1 & 61 & 1 & 73\\
				Ownership &1,639 & 10,465 & 1 & 886 & 1.0 &69.7 \\[2pt]
				\hline
		\end{tabular}}
	\end{center}
\end{table}

\begin{figure}[ht]
	\caption{Estimated block structure for the synthetic dataset}
	\label{fig:truth}
	\begin{subfigure}{.245\textwidth}
		\centering
		\includegraphics[width=.99\linewidth]{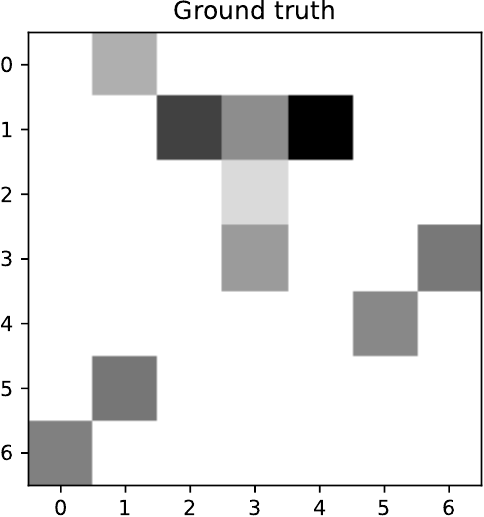}  
		\caption{Ground truth}
		\label{fig:sub-first}
	\end{subfigure}
	\begin{subfigure}{.245\textwidth}
		\centering
		\includegraphics[width=.99\linewidth]{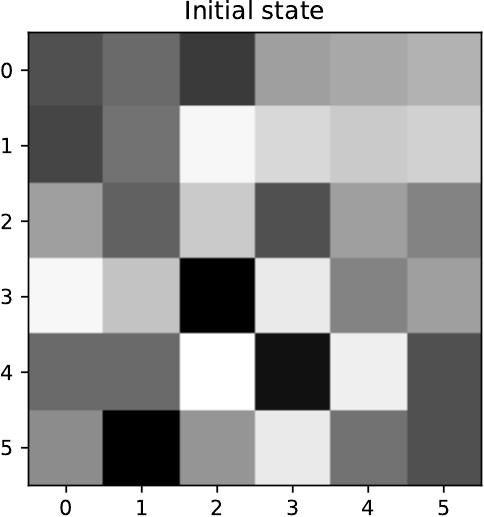}  
		\caption{Initial state}
		\label{fig:sub-first}
	\end{subfigure}
	\begin{subfigure}{.245\textwidth}
		\centering
		\includegraphics[width=.99\linewidth]{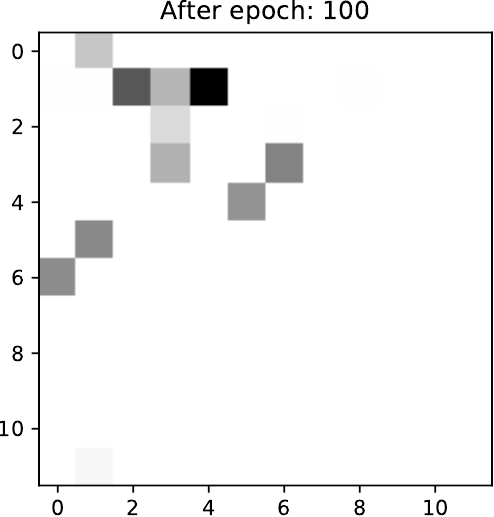}  
		\caption{100 epochs}
		\label{fig:sub-first}
	\end{subfigure}
	\begin{subfigure}{.245\textwidth}
		\centering
		\includegraphics[width=.99\linewidth]{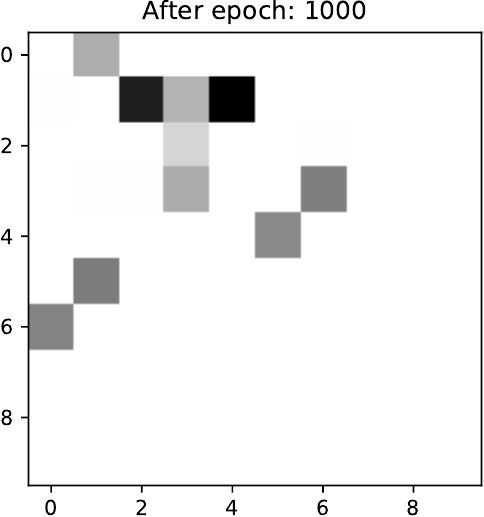}  
		\caption{1,000 epochs}
		\label{fig:sub-first}
	\end{subfigure}
\end{figure}

The motivation behind using a synthetic dataset is to illustrate whether our proposed algorithm recovers the ground truth block structure.   In this experiment, we used all the edges in the synthetic data for training. Fig.\ref{fig:truth} shows the result of running the algorithm for 1,000 epochs\footnote{One epoch comprises sampling all the edges in the training example once.}.  We confirm that after 100 epochs, the algorithm almost found the right block structure, and after 1,000 epochs, the result became more stable.  Thus, we conclude that our model correctly uncovers the latent block structure of a given network.

\subsection{Quantitative comparison}

We compared the performance of NDMDND with that of five models: SBM\cite{Holland1981,Peixoto2017}, degree corrected SBM\cite{Karrer2011,Peixoto2017}, weighted SBM\cite{Peixoto2018}, nested SBM\cite{Peixoto2013}, and MDND\cite{Williamson2016}.  For SBM-type models, we used the state-of-the-art graph tool library\cite{graph-tool2014}, which uses the minimum description length principle to determine the number of blocks used in the SBM.  Hence, it can be considered as a competitive alternative to the infinite relation model\cite{Kemp2006}.  The degree corrected SBM further takes into account the heterogeneous degree distribution of nodes.  For the weight function in the weighted SBM, we used the lognormal distribution for the synthetic data and an exponential distribution for the stock ownership data\footnote{We also tried the lognormal distribution for the stock ownership data, but it resulted in inferior performance.}. The nested SBM is a further extension of the SBM, which considers the fact that blocks themselves form a higher-level block structure.  This additional layer may enhance the predictive probability of an unseen edge by taking into account the nodes that may be softly classified into multiple groups, akin to the mixed membership SBM\cite{Airoldi2008}.

We used a link prediction task as our basis for quantitative comparison.  For both datasets, we used 80 percent of the data (i.e., edge list) as our training dataset and the remainder as our test dataset.  We trained all our models using the training dataset and measured the model's performance using the test dataset.  The models that we compared have different likelihood functions. Some can even model edge weights. Hence, we compared the models using a simple binary classification task.  For the stock ownership data, evaluating all the negative edges took so much time that it was impossible to assess the SBM models' performance.  Hence, we sampled $100,000$ negative edges instead of using all the negative samples. A standard metric used in binary classification is the area under the receiver operating curve (AUC-ROC).  However, the AUC-ROC overestimates the performance when the dataset is highly imbalanced, which applies to link prediction \cite{Yang2014}.  Moreover, theoretically, a model can only outperform in terms of the AUC-ROC when it outperforms in terms of the area under the precision-recall curve (AUC-PR)\cite{Davis2006}. Therefore, we used the AUC-PR score for the primary comparison.  Despite this, we also reported the AUC-ROC scores.

\begin{table}
	\caption{Predictive performance}
	\label{predictive}
	\begin{center}
		\resizebox{0.7\columnwidth}{!}{
			\begin{tabular}{c@{\quad}cccc}
				\hline
				& Synthetic &  & Ownership & \\
				Model &AUC-ROC& AUC-PR& AUC-ROC& AUC-PR\\[2pt]
				\hline\rule{0pt}{12pt}
				SBM & 0.956 & 0.414 & 0.966 & 0.575 \\
				DCSBM & 0.963 & 0.364 & 0.971 & 0.583 \\
				Nested SBM & 0.969 & 0.412 & \textbf{0.974} & 0.599 \\
				Weighted DCSBM & 0.971 & 0.672 & 0.97 & 0.568 \\
				MDND & 0.918 & 0.298 & 0.893 & 0.477 \\
				NDMDND & \textbf{0.983} & \textbf{0.736} & 0.968 & \textbf{0.673} \\[2pt]
				\hline
		\end{tabular}}
	\end{center}
\end{table}

Table \ref{predictive} summarizes the results.  It shows that NDMDND outperformed in terms of the AUC-PR quite significantly on both datasets.  In terms of the AUC-ROC, all the models' performance was almost the same,  except for MDND, which was significantly inferior on both datasets.  This inferior performance quite clearly highlights the limitations of the simple diagonal block structure, highlighting the importance of using our proposed NDMDND.

\subsection{Estimated block structure}
\begin{figure}
	\begin{subfigure}{.4\textwidth}
		\caption{Estimated block structure of stock ownership}
		\label{fig:stock}
		\centering
		\includegraphics[width=.99\linewidth]{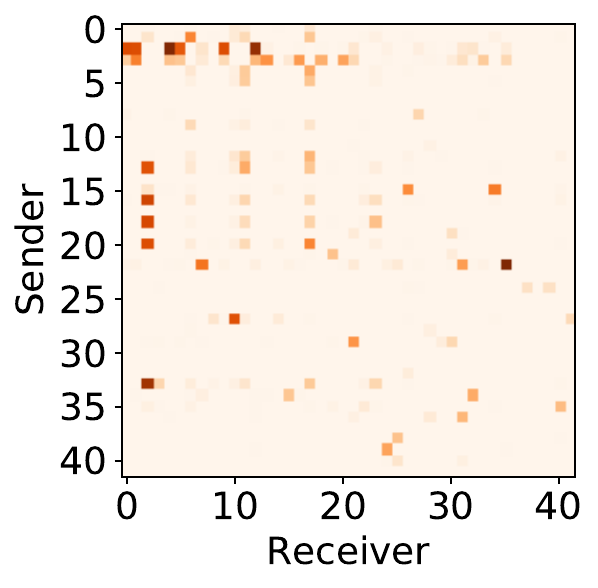}  
	\end{subfigure}
	\begin{subfigure}{.54\textwidth}
		\caption{Most probable nodes in a given block}
		\label{nodeid}
		\begin{center}
			\resizebox{1.\columnwidth}{!}{
				\begin{tabular}{c@{\quad}c}
					\hline
					Block & Nodes \\[2pt]
					\hline\rule{0pt}{12pt}
					2 & BlackRock, Fidelity Investments,\\
					& State Street Global Advisors\\
					3 & Vanguard Group, BlackRock,\\
					& Royce and Associates\\
					7 & Vanguard Group, Norges Bank,\\
					& Legal and General Investment Management\\
					10 & SPDR fund, iShares Morningstar\\
					& Permian Basin Royalty Trust\\
					15 & Vanguard group, TD Asset Management,\\
					& Brookfield Asset Management\\
					22 & Norges Bank, Schroder Investment Management,\\
					& Legal and General Investment Management\\
					26 & Manulife Financial, Transalta\\
					& Canadian Natural Resources\\
					27 & Morgan Stanley Wealth Management, SPDR fund,\\
					& Permian Basin Royalty Trust\\
					34 & TD Asset Management, TD Securities,\\
					& Investment Group Wealth Management\\
					35 & Vanguard Group, State Street Corporation\\
					& Legal and General Investment Management\\[2pt]
					\hline
			\end{tabular}}
		\end{center}
	\end{subfigure}
\end{figure}

Fig.\ref{fig:stock} shows the estimated block structure for the stock ownership data.  First, just by looking at the block structure matrix, we can see that several blocks are responsible for holding many of the other stocks in the dataset.  The most prominent blocks are blocks 2 and 3, which hold many stocks in the dataset.  As shown in Table \ref{nodeid}, these two blocks include companies such as ``BlackRock, Fidelity Investments, State Street Global Advisor,'' which are famous global asset management companies.  Another block pair that is quite huge in terms of the number of edges is block 22 to 35. Block 22 mainly contains European companies, whereas block 35 is a mixture of Canadian, U.S., and European asset managers.  Another interesting block is block 26, which contains mostly Canadian companies owned by block 15.  Block 15 is also mainly comprised of Canadian companies. Finally, block 10 does not own any stocks but is owned by many other nodes.  This is not surprising because block 10 mainly comprises exchange-traded funds.

\section{Conclusion}

In this paper, we proposed an edge exchangeable block model that estimates the latent block structure of complex networks.  Because the model is edge exchangeable, it reproduces the sparsity and heavy-tailed degree distribution that its random graph counterpart (i.e., SBM) fails to consider.  We tested our model using one synthetic dataset and one real-world stock ownership dataset and showed that our model outperformed state-of-the-art models.

\section{Acknowledgment}
R.H. was supported by Grant-in-Aid for Young Scientists \#20K20130, JSPS.  T.O. was supported by Grant-in-Aid for Scientific Research (A) \#19H01114, JSPS.  We thank Maxine Garcia from Edanz Group for editing a draft of this manuscript.

\bibliographystyle{unsrt}
\bibliography{reference}
\end{document}